# Recurrent Optical Spectrum Slicers as multi-λ processors for WDM optical equalization of IM/DD channels

Kostas Sozos [1], Francesco Da Ros [2], George Sarantoglou [3], Charis Mesaritakis [3], Adonis Bogris [1]

[1] Department of Informatics and Computer Engineering, University of West Attica, Aghiou Spiridonos, 12243, Egaleo, Athens, Greece, ksozos@uniwa.gr, abogris@uniwa.gr
[2] Department of Electrical and Photonics Engineering, Technical University of Denmark, Ørsteds Plads, DK-2800 Kgs. Lyngby, Denmark (fdro@dtu.dk)
[3] Department of Biomedical Engineering, University of West Attica, Aghiou Spiridonos, 12243, Egaleo, Athens, Greece, cmesar@uniwa.gr

**Abstract** *We report recurrent optical spectrum slicers as multi-wavelength optical processors in dispersion impaired IM/DD transmission systems with the use of programmable photonics. We demonstrate simultaneous equalization of three 64 Gb/s PAM-4 wavelengths after 75 km in the C-band and high tolerance to wavelength grid.* ©2025 The Authors

**Introduction**
Recent advances in high-speed transmission systems have brought 100 Gbaud technology in the forefront of 800G connectivity [1]. This evolution happened almost in parallel with the first 200 Gbaud demonstrations [2]. The digital equalisation of such fast links is mainly accomplished with sophisticated and complex equalisers like maximum-likelihood sequence estimators [1]. Apart from severely raising the power consumption, these equalisers are dispersion limited to 5-10 km links for the wavelengths in the edges of the O-band, necessitating denser grids than the typical one of 10 nm [2]. On top of that, they remain tailored to the specific channel parameters that were trained for, such as the transmission distance, the channel-related chromatic dispersion, the wavelength division multiplexing (WDM) grid etc. Their retraining under new transmission conditions requires higher effort in comparison to the straightforward Feed-Forward equalisers (FFE). Multi-channel equalisation based on simple FFE, agnostic to the channel spacing, the modulation format and the dispersion effect, i.e. independent on the transmission distance, the baudrate and the channel wavelength, can offer significant gains in both flexibility and digital processing complexity. Optical pre-processing aims to be an attractive alternative to the inherent limitations of digital signal processing (DSP), by equalising dispersive links while keeping the power consumption in reasonable levels [3]. However, by exploiting multiple extra optical components, optical equalisers tend to increase the power budget, the footprint and the cost, which are already limited for pluggable 800G transceivers.

Very recently, we proposed and experimentally demonstrated recurrent optical spectrum slicing (ROSS) receivers as potential optical accelerators in mitigating transmission impairments for both intensity modulated and coherently modulated signals [4-5]. We showcased unique compensation capabilities and detection performance for both 32 Gbaud PAM-4 and 16-QAM signals using two ROSS nodes, direct detection in two photoreceivers and digital FFE for combatting dispersion over 50 km in the C-band. Both demonstrations considered single wavelength transmission.

In this paper, we extend the previous work by experimentally validating ROSS receivers as processors of wavelength division multiplexed (WDM) signals. By employing a single pair of ROSS nodes we simultaneously equalise three channels of 32 Gbaud PAM-4 each with approximately 200 GHz channel spacing. By varying the transmission length from 25 km to 75 km and the channel spacing from 193 to 215 GHz, we provide evidence that the two-node ROSS receiver can achieve satisfactory equalization performance for all the three channels and all the transmission scenarios. In this respect, we offer a solution of a minimal multi-channel optical accelerator which is agnostic to the modulation format and the accumulated dispersion as also in the channel spacing. As in [4-5], the ROSS nodes are implemented with the use of a reconfigurable photonic platform (iPRONICS, Smartlight) [6]. The successful equalisation of channels at 75 km in the C-band renders our proposition suitable for the equivalent, in terms of dispersion, 200 Gbaud links operating in the O-band for up to 15 km.

**WDM ROSS experimental setup**
Conceptually, the ROSS system consists of two recurrent filter nodes, followed by separate photodetectors [4]. Each node contains an optical filter, in our case a Mach-Zehnder delayed interferometer (MZDI) and an external feedback loop controlled with a phase shifter. The phase difference, ΔΦ, between the arms of the unbalanced MZDI, defines the central frequency and the shape of the filter transfer function, while the external phase shift can further modify this shape in a beneficial manner [5]. Here, we take advantage of the programmable unit cells (PUCs) of the Smartlight platform for implementing a configuration incorporating a MZDI with 4 PUCs path difference (ΔT=45 ps) as the optical filter and a feedback loop of 2 PUCs (length 1.622 mm and delay 22.5 ps). ΔT determines the free spectral range of the filter and the subsequent bandwidth, which are 22.2 GHz and 11.1 GHz respectively.

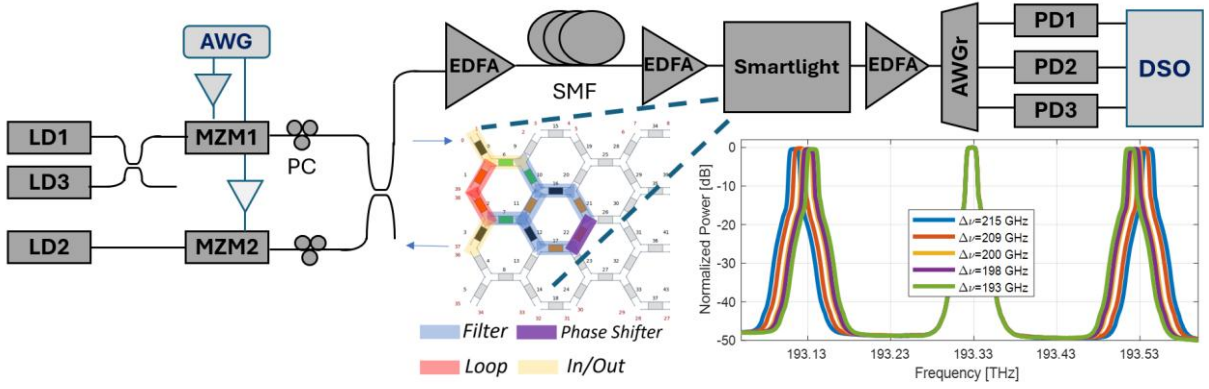

**Fig. 1:** The experimental setup employed in this work. Three WDM channels are equalised by combining two outputs of a single ROSS node consisting of a Mach-Zehnder delay interferometer in a loop. The node is implemented in the Smartlight platform as shown in the inset, while the channel spacing is varied from 193 to 215 GHz and the transmission distance from 25 km to 75 km. The launched power to the fiber is also varied with the addition of an extra EDFA prior to transmission from 0.8 dBm to 16d dBm.

The experimental setup is shown in Fig 1. At the transmitter side, a 3-channel WDM signal is generated using two Mach-Zehnder modulators (MZMs) driven by 32-GBd PAM4 sequences from a 64-GSa/s arbitrary waveform generator (AWG). The central channel ($f_{ch2}$ = 193.4 THz) originates from a narrow-linewidth fiber laser, while the two side channels are external cavity lasers (ECLs) and share the same modulator. After combining the three WDM channels in a 3-dB coupler, they are transmitted through a single-mode fibers (SMFs) of total length L=25, 50, and 75 km, for a total launch power of 0.8 dBm. At the output of the fiber, an Erbium-doped fiber amplifier (EDFA) pre-compensates for the high insertion loss (>20 dB) of the programmable integrated circuit used to implement the ROSS filter. Due to the polarization sensitivity of the PIC, polarization is manually aligned at the transmitter through polarization controllers (PCs). At the output of the ROSS filter, a second EDFA amplifies the signals, which are subsequently separated by a 200-GHz array waveguide grating (AWGr) demux and independently detected with three 50-GHz photodetectors (PDs) connected to three channels of an 80-GSa/s digital storage oscilloscope (DSO). We simultaneously measure the three channel outputs with the use of a single photonic node and then we vary the ΔΦ value, remeasuring sequentially for ΔΦ from 0 to 2π rad. Finally, we combine different pairs of node outputs, followed by a 31-tap T-spaced FFE, for equalizing each channel and improve bit-error ratio (BER) performance.

**Results**

We measured a wide number of link cases, in terms of channel spacing and transmission distance. The transmitted power was set to 0.8 dBm. Our purpose is to demonstrate how the same pair of ROSS nodes can be used for simultaneous decoding of multiple channels in a WDM system, as also to confirm the tolerance to changes of the basic link parameters, such as the dispersion or the wavelength. The optimum point for each case slightly changes, as it has been demonstrated in previous works [4]. Nevertheless, there can be found regimes of operation where the performance is satisfactory even when the accumulated dispersion triples (from 25 km to 75 km), or the grid changes from 193 GHz to 215 GHz. In Fig. 2, we showcase the optimization process followed by scanning all combinations of unique pairs of filters (125 measurements for 2π rad phase shift). We indicatively present the optimization process for all channels at 50 km distance (fig. 3a-c) and channel 2 behavior for two different grids at 50 km and 200 GHz grid at 75 km distance. Many regions of operation offering BER below the forward error correction (FEC) limit can be observed, but the region around the 0.5 rad and the 3 rad exhibits remarkable robustness to transmission conditions. This 2.5 rad phase difference between the two filters settings translates to 8.75 GHz frequency detuning between them. In Fig. 3, we illustrate how the three channels are simultaneously equalized for different grid values at 50 km showcasing a remarkable improvement compared to a direction detection system using a digital FFE (BER~7x10$^{-2}$) and even a nonlinear equalized based on a feedforward neural network (BER>4x10$^{-2}$). Notice the chosen grid values span both integer and non-integer multiples of the filter FSR. Better performance is attained at 200 GHz grid where all channels exhibit BER < $10^{-2}$. The worse BER performance observed in channels 1,3 (~$10^{-2}$) is attributed to their lower signal to noise ratio at the transmitter, due to the shared modulator, as well as due to spectral tilt in the EDFAs and slight offset filtering in the AWGr for non-200 GHz grids. All channels are affected by ~22 dB of losses induced by SmartLight. Through digital averaging we could observe significant BER improvement [7]. Digital averaging of 5 traces improves the BER performance over 4 times which proves that our setup is noise limited rather than dispersion limited. In fig. 5 the BER as a function of the distance for the three channels is depicted at 200 GHz grid. The superiority of ch2 is verified in all studied distances and BER < $2 \times 10^{-2}$ is provided for all WDM channels simultaneously up to 75 km of transmission.

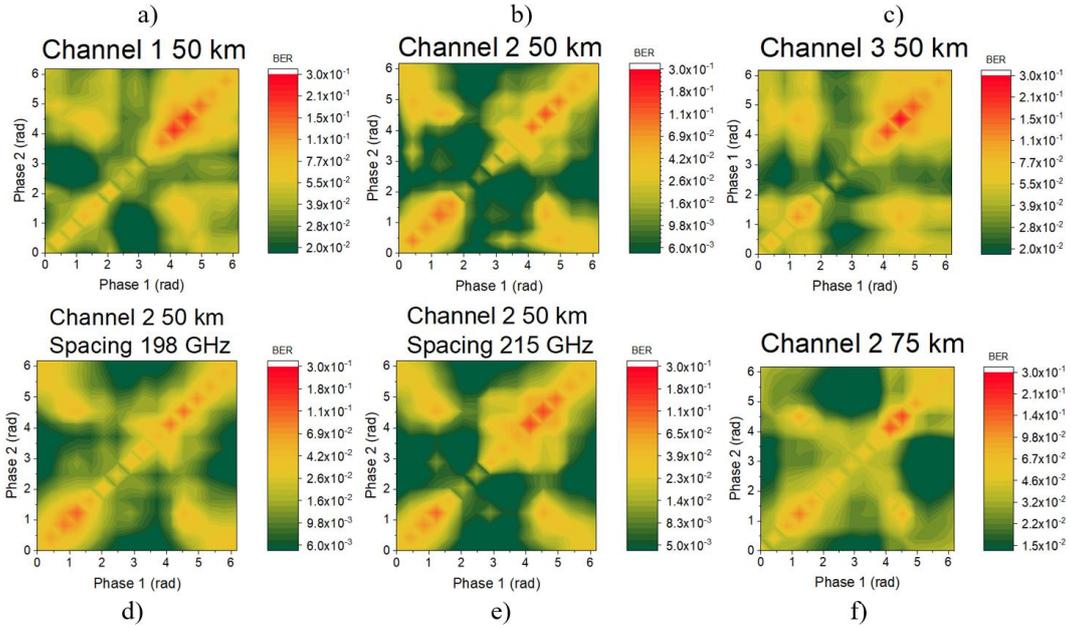

**Fig. 2:** a-c) Optimization of the three channels for the 50 km case with 200 GHz channel spacing. In all three cases, the best results lie in the same region around 0.5 rad and 3 rad, corresponding to a frequency detuning of about 8.75 GHz between the two filters. d-f) The optimization of channel 2 for two different grids at 50 km and at 75 km transmission distance (200 GHz grid). Fig. 3d-f) confirm that the same pair of filters can offer near-optimal performance for diverse

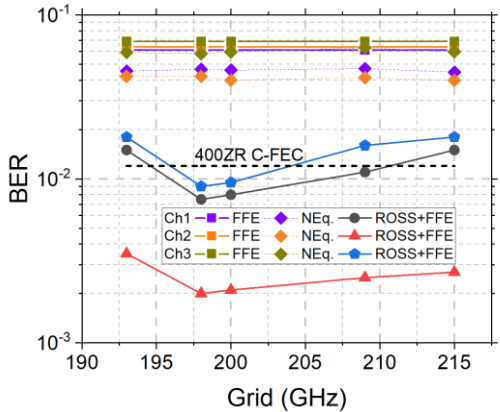

**Fig. 4:** The BER performance of the three channels with respect to grid separation and with comparison to simple direct detection followed by FFE at 50 km.

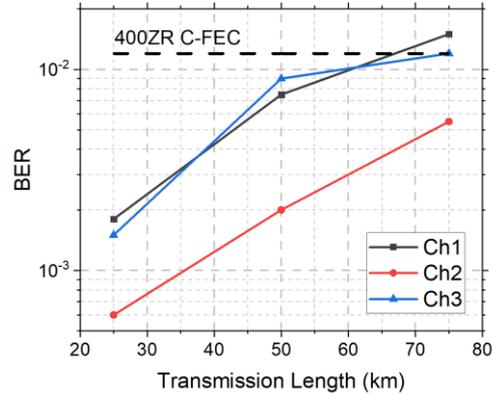

**Fig. 5:** The BER performance of the three channels as a function of distance at 200 GHz grid.

## Conclusions

In this work, we demonstrated that a two-node ROSS receiver can efficiently act as a multi-λ optical equaliser in dispersion impaired WDM links. We showcased the versatility of the optical pre-processor with respect to the transmission distance (up to 75 km in the C-band) and the WDM grid achieving BER < $2 \times 10^{-2}$ for all transmitted 32-Gbaud PAM-4 channels whilst a simple FFE after direct detection without ROSS nodes is limited to BER~$10^{-1}$. The BER performance can be further enhanced using custom designed ROSS filters with lower losses.

**Acknowledgements**

This work has been supported by Horizon Europe project PROMETHEUS under grant agreement 101070195, and the Villum Foundations project OPTIC-AI (grant n. VIL29344).